\documentclass[reprint,pre,amsmath]{revtex4-1}
\usepackage[T1]{fontenc}
\usepackage{array}
\usepackage{mathrsfs}
\usepackage{amsmath}
\usepackage{amsthm}
\usepackage{amssymb}
\usepackage{bm}
\usepackage{physics}
\usepackage{stmaryrd}
\usepackage{graphicx}
\usepackage{color}
\usepackage[colorlinks,linkcolor=blue,citecolor=blue]{hyperref}
\usepackage{newtxtext}
\usepackage[autostyle]{csquotes}
\usepackage{subfig}

\newcommand{\e}{\mathrm{e}}

\renewcommand{\log}{\operatorname{log}}
\MakeOuterQuote{"}

\begin{document}
	
	\title{Realizability of Heat Released in Nondeterministic Processes Derived from a Fluctuation Theorem}
	\author{Jihai Zhu}
	\affiliation{Department of Physics and Astronomy, Rice University, Houston, Texas 77005, USA}
	\begin{abstract}
		We derive a condition for the realizability of heat released in nondeterministic processes from a fluctuation theorem, showing that this condition is strictly stronger than the probabilistic form of the second law of thermodynamics. With this realizability condition, the (generalized) Zurek's bound can be straightforwardly obtained and applied to a wider range of systems.
		
	\end{abstract}
	
	\maketitle

	\section{Introduction}
		The 2nd law of thermodynamics, in its \textit{classical form}:
		\begin{equation}
			\Delta S\ge0,\label{eq: 2nd law classical}
		\end{equation}
		has long been regarded as holding "the supreme position among the laws of Nature"\cite{eddingtonNaturePhysicalWorld1928} and "will never be overthrown"\cite{schilppAlbertEinsteinPhilosopherscientist1970}, for its elegant simplicity and vast applicability. While Loschmidt's paradox, formulated in 1876, presented an early challenge to the validity of the 2nd law, it wasn't until the 1990s, with the discovery of fluctuation theorems, that its limitations became apparent.
		
		Fluctuation theorems are equalities for the probability distribution functions for the thermodynamic quantities. First discovered in a thermostatted system in 1993\cite{evansProbabilitySecondLaw1993}, different fluctuation theorems have been discovered in various systems, thermostatted\cite{evansProbabilitySecondLaw1993,evansEquilibriumMicrostatesWhich1994a} or Hamiltonian\cite{jarzynskiNonequilibriumEqualityFree1997a,jarzynskiHamiltonianDerivationDetailed2000}, deterministic\cite{gallavottiDynamicalEnsemblesNonequilibrium1995} or stochastic\cite{crooksNonequilibriumMeasurementsFree1998,kurchanFluctuationTheoremStochastic1998}, and have been justified in many experiments. A key insight provided by the fluctuation theorems is that there exists a non-zero probability of the 2nd law being violated in non-equilibrium processes occurring in small systems. As a result, the classical form (\ref{eq: 2nd law classical}) of the 2nd law is incompatible with the fluctuation theorems. Fortunately, from these theorems a \textit{probabilistic form} of the 2nd law can be derived:
		\begin{equation}
			\Sigma(p_X)\equiv-S(p_X)+S(p_Y)+\langle Q\rangle_{p_X}\ge0,\forall p_X\in\Delta_{\mathcal{X}}.\label{eq: 2nd law probabilistic}
		\end{equation} 
		To explain (\ref{eq: 2nd law probabilistic}): suppose a system, coupled to a reservoir, is prepared in an initial distribution $p_X$ over initial states $x\in\mathcal{X}$, and it is driven to a final distribution $p_Y$ over final states $y\in\mathcal{Y}$. $S(p_X)$ and $S(p_Y)$ represent the Shannon entropy of $p_X$ and $p_Y$, while $\langle Q\rangle_{p_X}$ denotes the average amount heat released to the reservoir during the process, under the initial distribution $p_X$. We set $k_BT=1$ for simplicity. Thus, the probablistic form of the 2nd law states that, for any \textbf{physically realizable} process, the increase in Shannon entropy, plus the average heat released, must be non-negetive for any initial distribution of the system. 
		
		However, the probabilistic form of the 2nd law does not carry the same strength as fluctuation theorems, as it only considers the average behavior of a process, thereby neglecting the detailed information contained in the probabilistic nature of thermodynamic quantities. Various thermodynamic bounds which are stronger than (the probabilistic form of) the 2nd law have been proposed, such as the thermodynamic uncertainty relations, speed limits, and bounds on work fluctuations\cite{horowitzThermodynamicUncertaintyRelations2020,voUnifiedThermodynamickineticUncertainty2022,yoshimuraThermodynamicUncertaintyRelation2021,vanvuThermodynamicUnificationOptimal2023,cavinaOptimalProcessesProbabilistic2016}. Yet, none of these bounds provides a direct visualization of the regions allowed by the 2nd law compared to those allowed by the fluctuation theorems.
		
		In this article, we provide such a visualization by addressing the following problem:
		
		\textit{Suppose that a process is applied on a system with an initial configuration $\mathcal{X}$ and brings it to a final configuration $\mathcal{Y}$ (both $\mathcal{X}$ and $\mathcal{Y}$ are sets of states). The process is carried out independent of the initial state $x\in\mathcal{X}$. $P(y|x)$ denotes the probability that the system ends up in the final state $y\in\mathcal{Y}$, given that the system starts in the initial state $x\in\mathcal{X}$ . $Q(x\to y)$ (called the \textbf{heat release function}) denotes the average heat released to the reservoir, given that the system starts in $x$ and ends up in $y$. What is the condition for the pair $(P(y|x),Q(x\to y))$, so that such a process is physically realizable (i.e., can be realized by some physical apparatus)?}
		
		We will show that the fluctuation theorems lead to a condition for realizability, which is strictly stronger than the requirement of (the probabilistic form of) the 2nd law. By plotting pairs $(P,Q)$ allowed by this realizability condition and those by the 2nd law, we visualize the relative strength of these two.
		
		The condition for realizability from the fluctuation theorem was previously derived for Hamiltonian systems in \cite{turgutRelationsEntropiesProduced2009}. In this article, we derive the same condition from a fluctuation theorem, thus generalizing its range of applicability.
		
		This article is organized as follows. In section \ref{sec: Fluctuation Theorems} we introduce a fluctuation theorem  which is used to derive our main result. In section \ref{sec: Main Result} we derive the condition for realizability of heat released in nondeterministic processes from the fluctuation theorem, and demonstrate its strength over the 2nd law. In section \ref{sec: Applications} we apply the result to a noisy 1-bit channel and the (generalized) Zurek's bound. Section \ref{sec: Conclusions} provides a brief conclusion.

	\section{Fluctuation Theorems}\label{sec: Fluctuation Theorems}
	
	In this section, we introduce the fluctuation theorem with which we derive the main result of this article in the next section.
	
	To begin with, let's consider a system which is coupled to a reservoir and driven under some control parameter $\lambda(t)$, $t\in[0,\tau]$. We call $\lambda(t)$ as the \textbf{forward process}. During the process, the system (stochastically) travels a trajectory $\Gamma(t)$ in its phase space. $\Gamma(0)\equiv x\in\mathcal{X}$ is the initial state and $\Gamma(\tau)\equiv y\in\mathcal{Y}$ the final state. Note that although in the above we have assumed that $\mathcal{X}$ and $\mathcal{Y}$ are sets of microscopic states in the phase space, all the equations and arguments in this article hold for the case where $\mathcal{X}$ and $\mathcal{Y}$ are sets of coarse-grained macroscopic states. 
	
	The \textbf{detailed fluctuation theorem for trajectories} is written as
	\begin{equation}
		\frac{P[\Gamma]}{P^*[\Gamma^*]}=\e^{\sigma[\Gamma]},\label{eq: detailed FT for trajectories}
	\end{equation}
	where $P[\Gamma]$ denotes the probability that the system travels the trajectory $\Gamma$ in the process, and $\sigma[\Gamma]$ denotes the entropy production along the trajectory $\Gamma$. As in many fluctuation theorems, we also consider the \textbf{reverse process}, that is, the process where the control parameter evolves in a time-reversing way: $\lambda^*(t)\equiv\lambda(\tau-t)$. $\Gamma^*$ represents the reverse trajectory of $\Gamma$:
	\begin{multline}
			\Gamma^*(t)\equiv(\Gamma(\tau-t))^*\\
			=(q(\tau-t),p(\tau-t))^*=(q(\tau-t),-p(\tau-t)),
	\end{multline}
	and $P^*$ represents the prbability distribution function of the reverse process.
	
	The entropy production along a trajectory can be decomposed as
	\begin{multline}
		\sigma[\Gamma]=s_{\tau}(\Gamma(\tau))-s_0(\Gamma(0))+q[\Gamma]\\
		=-\log P_{\tau}(\Gamma(\tau))+\log P_{0}(\Gamma(0))+q[\Gamma],\label{eq: trajectory entropy}
	\end{multline}
	where $s_t(\Gamma(t))\equiv-\log P_t(\Gamma(t))$ denotes the trajectory entropy, and $q[\Gamma]$ denotes the heat released along the trajectory $\Gamma$.
	
	Substitue $\sigma(\Gamma)$ in (\ref{eq: detailed FT for trajectories}) with (\ref{eq: trajectory entropy}), we can write
	\begin{equation}
		\frac{P[\Gamma|\Gamma(0)]}{P^*[\Gamma^*|\Gamma^*(\tau)]}=\e^{q[\Gamma]}.\label{eq: detailed FT for trajectories_heat}
	\end{equation}
	
	With the above preparations, we now define $P(y,q|x)$ as the conditional prbability, that the initial state $x$ ends up in the final state $y$ with $q$ heat released, in the forward process, and similar for $P^*(x^*,-q|y^*)$. We have
	\begin{equation}
		\begin{split}
			P(y,q|x)&=\sum_{\Gamma:\Gamma(0)=x,\Gamma(\tau)=y,q[\Gamma]=q}P[\Gamma|x]\\
			&=\sum_{\Gamma:\Gamma(0)=x,\Gamma(\tau)=y,q[\Gamma]=q}P^*[\Gamma^*|y^*]\e^q\\
			&=\sum_{\Gamma^*:\Gamma^*(0)=y^*,\Gamma^*(\tau)=x^*,q[\Gamma^*]=-q}P^*[\Gamma^*|y^*]\e^q\\
			&=P^*(x^*,-q|y^*)\e^q,\label{der: FT_differential}
		\end{split}
	\end{equation}
	where in the second line we used (\ref{eq: detailed FT for trajectories_heat}). Rearrange the terms in (\ref{der: FT_differential}), we can write
	\begin{equation}
		\frac{P(y,q|x)}{P^*(x^*,-q|y^*)}=\e^{q},\label{eq: differential FT}
	\end{equation}
	and we call this equation the \textbf{differential fluctuation theorem}\cite{jarzynskiHamiltonianDerivationDetailed2000}.
	
	In the above arguments, we have assumed that for every trajectory $\Gamma$,
	\begin{equation}
		P[\Gamma]=0\Rightarrow P^*[\Gamma^*]=0,\label{eq: stochastic reversiblility}
	\end{equation}
	which means that every trajectory $\Gamma$ is \textbf{stochastically reversible}. However, in some cases, such as free expansion, the above condition (\ref{eq: stochastic reversiblility}) may be violated, that is, there exists trajectory $\Gamma$ such that,
	\begin{equation}
		P[\Gamma]=0\text{ and }P^*[\Gamma^*]\ne0.
	\end{equation}
	We call such trajectories \textbf{absolutely irreversible}\cite{murashitaNonequilibriumEqualitiesAbsolutely2014}. Since equation (\ref{eq: detailed FT for trajectories_heat}) is only satisfied by stochastically reversible trajectories, we have to rewrite (\ref{der: FT_differential}) as
	\begin{equation}
		\begin{split}
			P(y,q|x)&=\sum_{\Gamma:\Gamma(0)=x,\Gamma(\tau)=y,q[\Gamma]=q}P[\Gamma|x]\\
			&=\sum_{\Gamma:\Gamma(0)=x,\Gamma(\tau)=y,q[\Gamma]=q,P[\Gamma]\ne0}P^*[\Gamma^*|y^*]\e^q\\
			&=\sum_{\Gamma^*:\Gamma^*(0)=y^*,\Gamma^*(\tau)=x^*,q[\Gamma^*]=-q,P[\Gamma]\ne0}P^*[\Gamma^*|y^*]\e^q\\
			&\le P^*(x^*,-q|y^*)\e^q,\label{der: FT_differential_absolute irreversibility}
		\end{split}
	\end{equation}
	and the differential fluctuation theorem with absolute irreversibility becomes
	\begin{equation}
		\frac{P(y,q|x)}{P^*(x^*,-q|y^*)}\le\e^{q}.\label{eq: differential FT_absolute irreversibility}
	\end{equation}

	\section{Main Result}\label{sec: Main Result}
	In this section, we derive the condition for the realizability of $(P(y|x),Q(x\to y))$ from the differential fluctuation theorem (\ref{eq: differential FT}) and (\ref{eq: differential FT_absolute irreversibility}). For any $y\in\mathcal{Y}$, we have the following equations:
	\begin{equation}
		\begin{split}
			\sum_{x\in\mathcal{X}}P(y|x)\e^{-Q(x\to y)}=&\sum_{x\in\mathcal{X}}P(y|x)\e^{-\langle q\rangle_{P(q|x,y)}}\\ 
			\le&\sum_{x\in\mathcal{X}}\sum_q P(y|x)P(q|x,y)\e^{-q}\\
			=&\sum_{x\in\mathcal{X}}\sum_q P(y,q|x)\e^{-q}\\
			=&\sum_{x\in\mathcal{X}}\sum_q P^*(x^*,-q|y^*)\\
			=&\sum_{x\in\mathcal{X}}P^*(x^*|y^*)\\
			=&1,
		\end{split}\label{der: heat condition_FT}
	\end{equation}
	where the second line follows from Jensen's inequality and the fourth line follows from (\ref{eq: differential FT}). We may substitute "=" with "$\le$" in the fourth line if absolute irreversibility is considered. Nevertheless, for any physically realizable $(P,Q)$, the following condition holds:
	\begin{equation}
		\sum_{x\in\mathcal{X}}P(y|x)\e^{-Q(x\to y)}\le1,\quad\forall y\in\mathcal{Y}.\label{eq: heat condition_FT}
	\end{equation}
	This condition is both necessary and sufficient (which is proved in \cite{turgutRelationsEntropiesProduced2009}), and is therefore equivalent to the fluctuation theorems. By "sufficient" we mean that, for any $(P,Q)$ which satisfies the condition, we can construct a physical apparatus (a Hamiltonian for the system and the reservoir) and a control protocol to carry out the process.
	
	To interprete this condition more physically, we observe from (\ref{der: heat condition_FT}) that the equality in (\ref{eq: heat condition_FT}) holds whenever the process on trajectories $P[\Gamma]$ is not absolutely irreversible and that the averaged term in Jensen's inequality is constant:
	\begin{equation}
		q[\Gamma]=Q(\Gamma(0)\to\Gamma(\tau)),\quad\forall\Gamma,
	\end{equation} 
	i.e., given the initial state $x$ and the final state $y$, any trajectory $\Gamma$ that travels from $x$ to $y$ releases the same amount of heat $q[\Gamma]=Q(x\to y)$.
	
\begin{figure}
		\centering
		\includegraphics[width=0.45\textwidth]{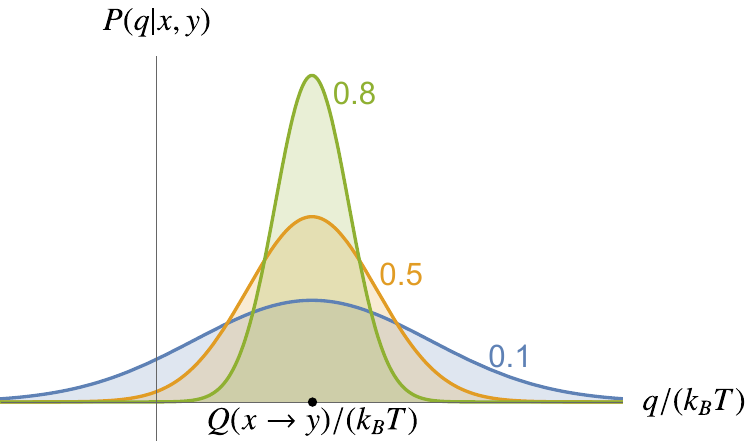}
		\caption{Distributions of heat release given the initial and final states. The distributions are assumed to be normal. Numbers labeled next to the curves are the sums in (\ref{eq: heat condition_FT})} for each distribution, calculated with (\ref{eq: heat distribution_summation}).
		\label{fig: Gaussian distribution}
	\end{figure}

	Quantitively, let's assume that the distribution of heat release given the initial and final state, $P(q|x,y)$, is a normal distribution centered at $Q(x\to y)$ (FIG.\ref{fig: Gaussian distribution}):
	\begin{equation}
		P(q|x,y)=\frac{1}{\sqrt{2\pi}\sigma}\e^{-\frac{(q-Q(x\to y))^2}{2\sigma^2}}.\label{eq: heat distribution}
	\end{equation}
	Apply (\ref{eq: heat distribution}) to (\ref{der: heat condition_FT}), we have
	\begin{equation}
		\sum_{x\in\mathcal{X}}P(y|x)\e^{-Q(x\to y)}=\frac{\e^{\langle-q\rangle_{P(q|x,y)}}}{\langle\e^{-q}\rangle_{P(q|x,y)}}=\e^{-\sigma^2/2},\label{eq: heat distribution_summation}
	\end{equation}
	that is, the narrower the distribution of heat release given the initial and final state is, the closer the sum in (\ref{eq: heat condition_FT}) gets to 1.

	\begin{figure}
		\centering
		\includegraphics[width=0.28\textwidth]{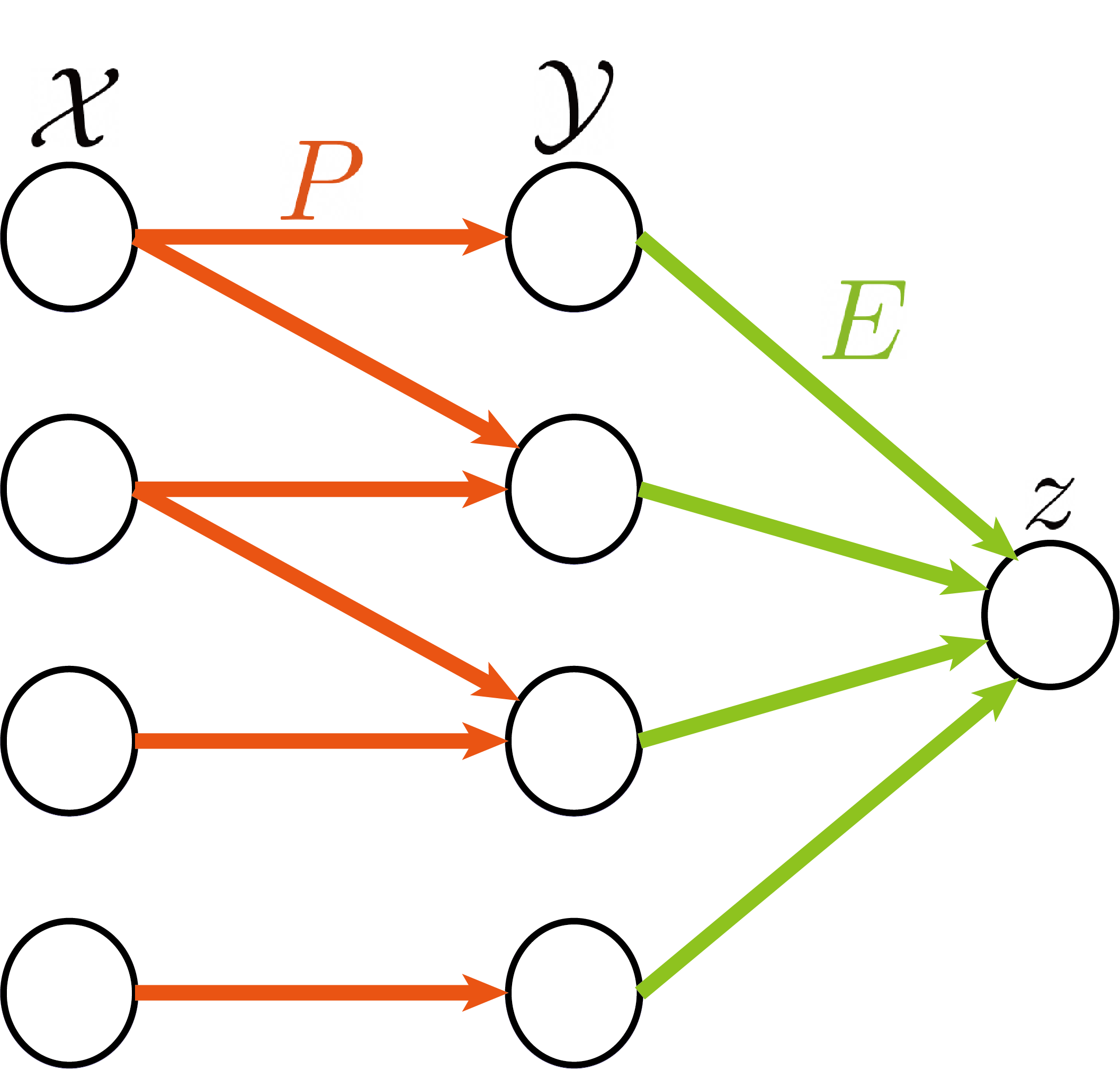}
		\caption{The composite process of $P(y|x)$ and the erasure process $E:y\in\mathcal{Y}\mapsto z$.}
		\label{fig: erasure surgery}
	\end{figure}
	
	To prove that condition (\ref{eq: heat condition_FT}) is stronger than (the probabilistic form of) the 2nd law (\ref{eq: 2nd law probabilistic}), we use a result from \cite{kolchinskyThermodynamicCostsTuring2020}. The result states that for any deterministic process $f:\mathcal{X}\to\mathcal{Y}$ and average heat release function $Q(x)\equiv\sum_{y\in\mathcal{Y}}P(y|x)Q(x\to y)$, the pair $(f,Q)$ satisfies the 2nd law (\ref{eq: 2nd law probabilistic}) iff
	\begin{equation}
		\sum_{x\in\mathcal{X}:f(x)=y}\e^{-Q(x)}\leq 1,\quad\forall y\in\mathcal{Y}.\label{eq: heat condition_deterministic}
	\end{equation}
	What's more, it is proved that the equality holds in (\ref{eq: heat condition_deterministic}) iff there exists $p_X\in\Delta_{\mathcal{X}}$ such that $\Sigma(p_X)=0$, and that $\Sigma(p_X)=0$ iff $Q(x)=-\log p_X(x)$. At this point, we note that the condition (\ref{eq: heat condition_FT}) recovers (\ref{eq: heat condition_deterministic}) when $P(y|x)$ can be identified to a deterministic process. This coincidence implies that there is no difference for fluctuation theorems and the 2nd law in deterministic processes, if only initial-state-averaged heat release is considered.
	
	To apply the above result for the nondeterministic process $P(y|x)$, we can apply an \textbf{erasure process} $E:y\in\mathcal{Y}\mapsto z$ (together with any realizable heat release function $Q'(y)$ of process $E$) after $P$, so that both $E$ and $E\circ P$ are deterministic processes (FIG.\ref{fig: erasure surgery}). 
	
	The 2nd law for process $P$ requires that
	\begin{equation}
		\Sigma_P(p_X)\ge0,\quad\forall p_X\in\Delta_{\mathcal{X}},\label{eq: 2nd law_P}
	\end{equation}
	while for process $E\circ P$ requires that
	\begin{equation}
		\Sigma_{E\circ P}(p_X)=\Sigma_P(p_X)+\Sigma_{E}(Pp_X)\ge0,\quad\forall p_X\in\Delta_{\mathcal{X}},
	\end{equation}
	Since for any initial distribution $p_X$, we can set
	\begin{equation}
		Q'(y)=-\log[Pp_X](y)=-\log\sum_{x\in\mathcal{X}}P(y|x)p_X(x),
	\end{equation}
	so that $\Sigma_E(Pp_X)=0$, condition (\ref{eq: 2nd law_P}) holds iff
	\begin{equation}
		\Sigma_{E\circ P}(p_X)\ge0,\quad\forall p_X\in\Delta_{\mathcal{X}},\forall\text{realizable }Q'(y).\label{eq: 2nd law_EP}
	\end{equation} 
	
	Apply (\ref{eq: heat condition_deterministic}) to $E\circ P$, we can write an equivalent form of (\ref{eq: 2nd law_EP}):
	\begin{equation}
		\sum_{x\in\mathcal{X}}\e^{-\sum_{y\in\mathcal{Y}}P(y|x)(Q(x\to y)-\log p_Y(y))}\le1,\forall p_Y\in\Delta_{\mathcal{Y}},\label{eq: heat condition_composition}
	\end{equation}
	 where we have assumed that $Q'(y)=-\log p_Y(y)$. We have
	 \begin{equation}
	 	\begin{split}
	 		&\sum_{x\in\mathcal{X}}\e^{-\sum_{y\in\mathcal{Y}}P(y|x)(Q(x\to y)-\log p_Y(y))}\\
	 		\le&\sum_{x\in\mathcal{X},y\in\mathcal{Y}}P(y|x)p_Y(y)\e^{-Q(x\to y)}\\
	 		\le&\sum_{y\in\mathcal{Y}}p_Y(y)\\
	 		=&1,\label{der: FT stronger than 2nd law}
	 	\end{split}
	 \end{equation}
	where the first inequality follows from Jensen's inequality, and the second inequality follows from assumption (\ref{eq: heat condition_FT}). Thus, we have proved that the condition (\ref{eq: heat condition_FT}) obtained from the fluctuation theorems is stronger than the 2nd law (\ref{eq: 2nd law probabilistic}).
	
	One might wonder whether there exist pairs $(P,Q)$ that are not only physically realizable (condition (\ref{eq: heat condition_FT})), but also dissipationless (i.e., where $\Sigma(p_X)=0$ for some $p_X\in\Delta_{\mathcal{X}}$). To answer this, note that $(P,Q)$ is dissipationless iff the equalities in (\ref{der: FT stronger than 2nd law}) hold, which requires
	\begin{equation}
		\sum_{x\in\mathcal{X}}P(y|x)\e^{-Q(x\to y)}=1,\quad\forall y\in\mathcal{Y},\label{eq: dissipationless condition_FT}
	\end{equation}
	and that there exists a function $F(x)$ such that
	\begin{equation}
		Q(x\to y)-\log p_Y(y)=F(x).\label{eq: dissipationless condition_Jensen}
	\end{equation}
	Applying (\ref{eq: dissipationless condition_Jensen}) to (\ref{eq: dissipationless condition_FT}), we obtain
	\begin{equation}
		p_Y(y)=\sum_{x\in\mathcal{X}}P(y|x)\e^{-F(x)}.\label{eq: dissipationless condition_pY}
	\end{equation}
	Summing up (\ref{eq: dissipationless condition_pY}) over $y\in\mathcal{Y}$, we have
	\begin{equation}
		\sum_{x\in\mathcal{X}}\e^{-F(x)}=1,
	\end{equation}
	which allows us to assume $F(x)=-\log p_X(x)$. Combining this with (\ref{eq: dissipationless condition_pY}), we obtain $p_Y=Pp_X$. Therefore, a physically realizable pair $(P,Q)$ is dissipationless iff there exists $p_X\in\Delta_{\mathcal{X}}$ such that
	\begin{equation}
		Q(x\to y)=-\log p_X(x)+\log [Pp_X](y).
	\end{equation}
	Given the process $P(y|x)$, the set of all the realizable heat release functions $Q(x\to y)$ forms an $(|\mathcal{X}|+|\mathcal{Y}|)$ dimensional manifold, with an $(|\mathcal{X}|+|\mathcal{Y}|-1)$ dimensional boundary, and the set of all the dissipationless heat release functions $Q(x\to y)$ forms an $(|\mathcal{X}|-1)$ dimensional submanifold on this boundary.

	\section{Applications}\label{sec: Applications}
	
	\subsection{Noisy 1-bit Channel}
\begin{figure}
		\centering
		\includegraphics[width=0.25\textwidth]{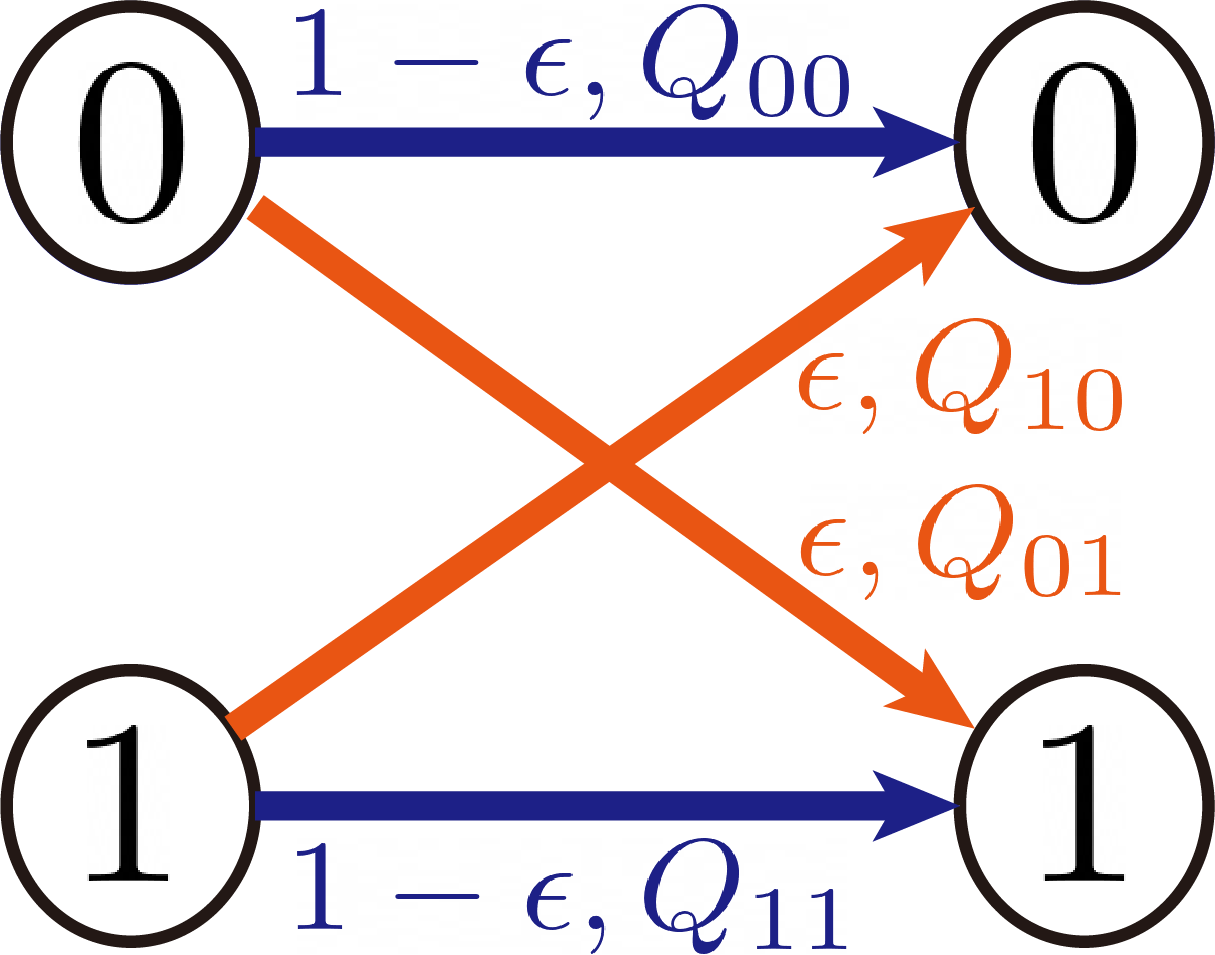}
		\caption{A noisy 1-bit channel.}
		\label{fig: noisy 1-bit channel}
	\end{figure}

	Here we illustrate the above results with a simple model of nondeterministic processes: a noisy 1-bit channel\cite{mackayInformationTheoryInference2003} (FIG.\ref{fig: noisy 1-bit channel}).
	
	The noisy 1-bit channel is a process transmitting one bit of information in a 2-state system with $\mathcal{X}=\mathcal{Y}=\{0,1\}$, with an \textbf{error probability} of $\epsilon$, that is, $P(0|0)=P(1|1)=1-\epsilon$ and $P(1|0)=P(0|1)=\epsilon$. We assume that the "noise" comes from the channel being coupled to a heat reservoir, and distinct amount of heat is released to the reservoir along with each correct or error transfer: $Q_{ij}\equiv Q(i\to j),i,j=0,1$.
	
\begin{figure}
		\centering
		\subfloat[$\epsilon=0.1$]{
			\label{sfig:example-fig-logo-fig}
			\includegraphics[width=0.2\textwidth]{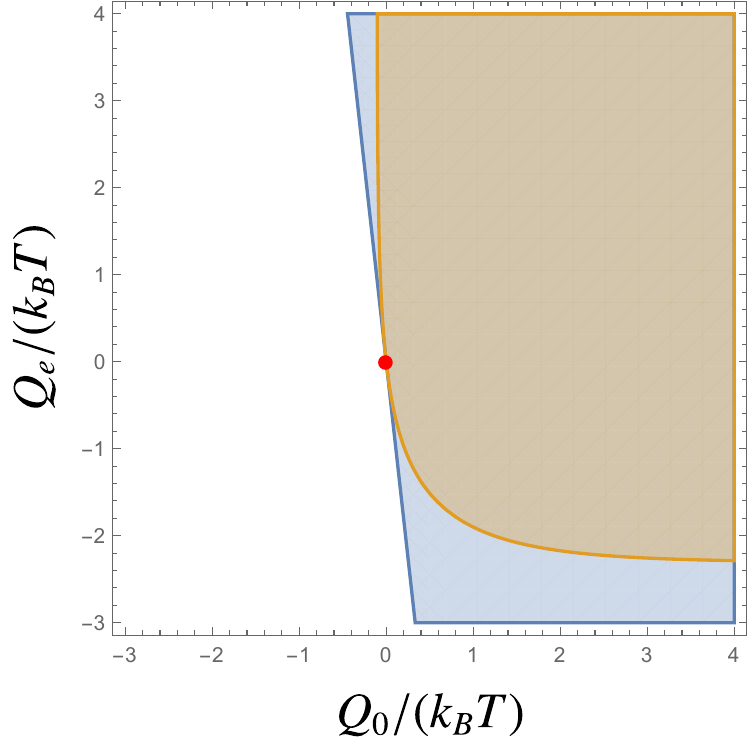}}\hspace{1em}
		\subfloat[$\epsilon=0.3$]{
			\includegraphics[width=0.2\textwidth]{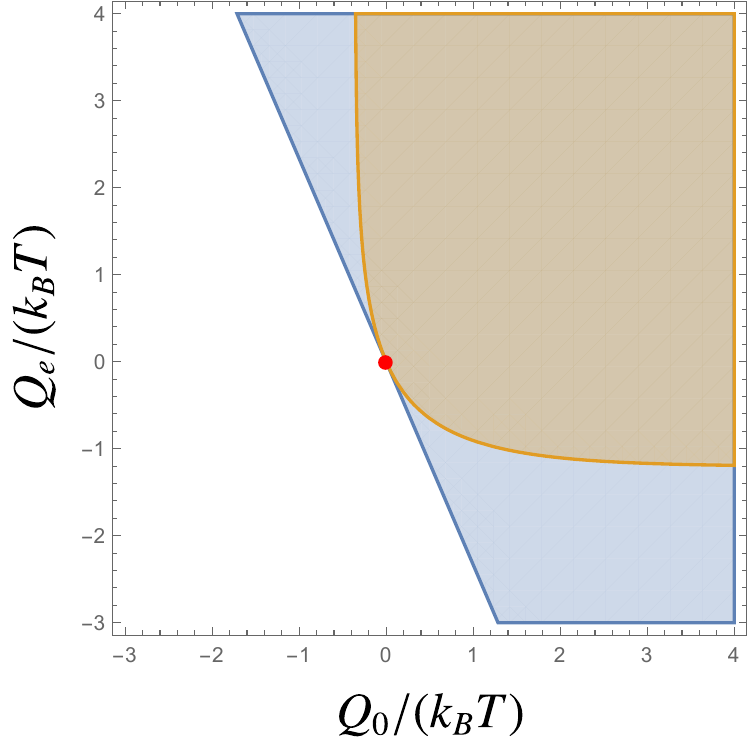}}\hspace{0em}
		\subfloat[$\epsilon=0.5$]{
			\includegraphics[width=0.2\textwidth]{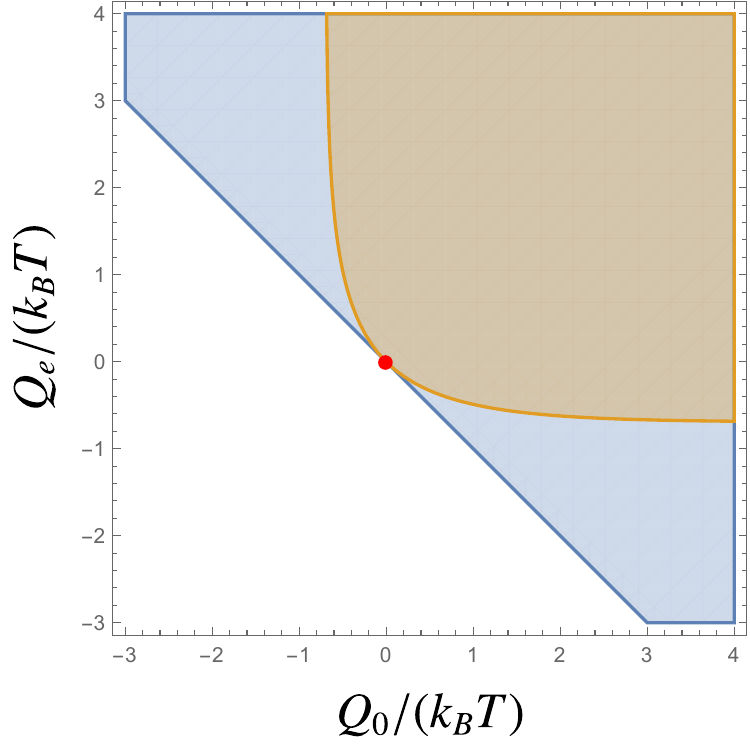}}\hspace{1em}
		\subfloat[$\epsilon=0.8$]{
			\includegraphics[width=0.2\textwidth]{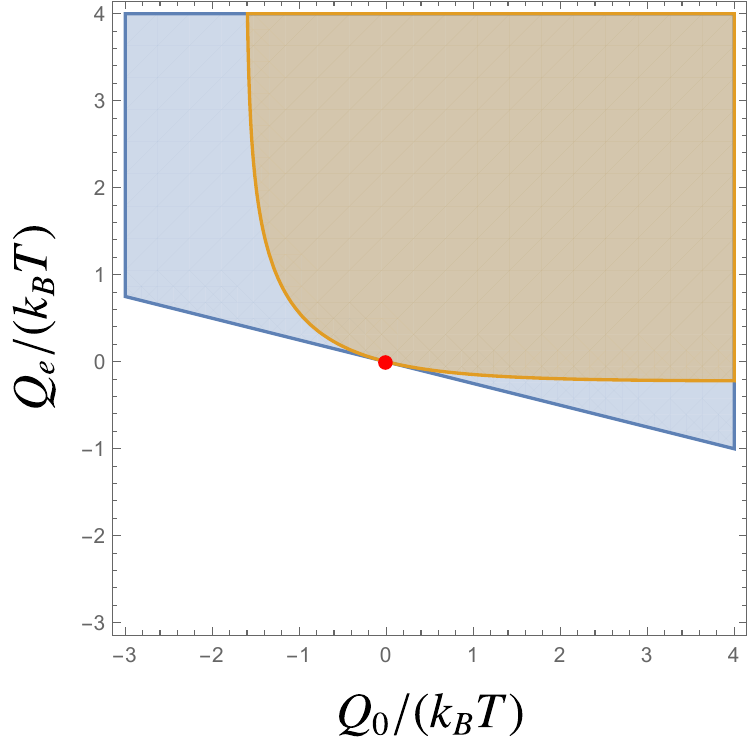}}
		\caption{$(Q_0,Q_e)$ that that is realizable (condition (\ref{eq: noisy 1-bit channel_sym_condition_FT})) is plotted in orange, and that satisfies the 2nd law (condition (\ref{eq: noisy 1-bit channel_sym_condition_2nd law})) is plotted in blue, for error probability $\epsilon=0.1,0.3,0.5,0.8$. The two regions intersect at $(0,0)$ (red dot).}
		\label{fig: comparison}
	\end{figure}

	First, let's consider the case where $Q_{00}=Q_{11}=Q_0$ and $Q_{01}=Q_{10}=Q_e$ (i.e., the channel is symmetric). The condition for realizability (\ref{eq: heat condition_FT}) requires that
	\begin{equation}
		(1-\epsilon)\e^{-Q_0}+\epsilon\e^{-Q_e}\le1,\label{eq: noisy 1-bit channel_sym_condition_FT}
	\end{equation}
	while the 2nd law (\ref{eq: 2nd law probabilistic}) requires that
	\begin{equation}
		(1-\epsilon)Q_0+\epsilon Q_e\ge0.\label{eq: noisy 1-bit channel_sym_condition_2nd law}
	\end{equation}
	By plotting $(Q_0,Q_e)$ that satisfies inequalities (\ref{eq: noisy 1-bit channel_sym_condition_FT}) and (\ref{eq: noisy 1-bit channel_sym_condition_2nd law}) (FIG.\ref{fig: comparison}) for different error probability $\epsilon$, we clearly see that the realizability condition (\ref{eq: heat condition_FT}), which is equivalent to the differential fluctuation theorem, is strictly stronger than the 2nd law. We note that for symmetric channels, there is only one pair of realizable $(Q_0,Q_e)=(0,0)$ that is dissipationless.
	
\begin{figure}
		\centering
		\includegraphics[width=0.4\textwidth]{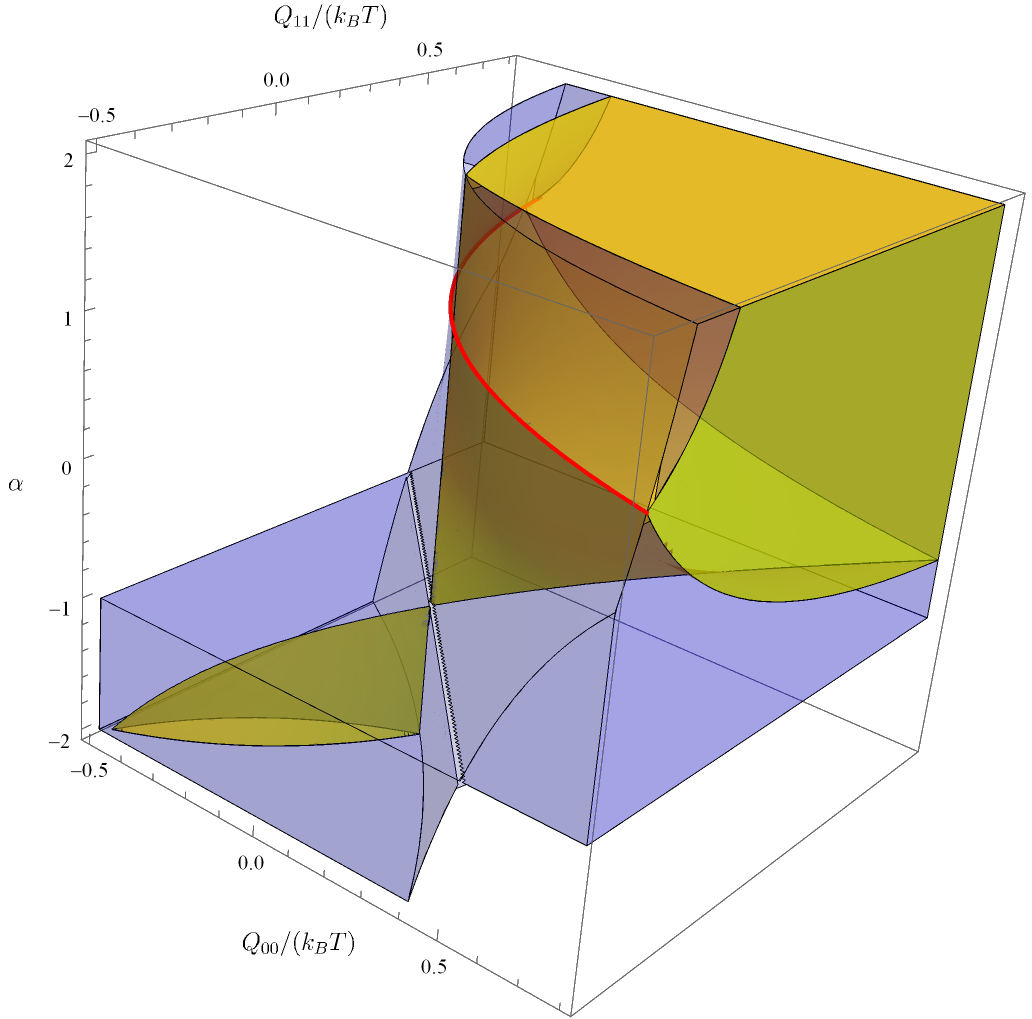}
		\caption{$(Q_{00},Q_{11},\alpha)$ that is physically realizable (\ref{eq: noisy 1-bit channel_asym_condition_FT}) is plotted in yellow, and that satiesfies the 2nd law (\ref{eq: noisy 1-bit channel_asym_condition_2nd law}) is plotted in blue. The red curve shows the intersection of the two regions.}
		\label{fig: coincidence curve}
	\end{figure}

	Next, let's consider the case where $Q_{01}/Q_{00}=Q_{10}/Q_{11}=\alpha$. For simplicity of analysis, we assume $\epsilon=1/2$. Similarly, we have the following condition for realizability:
	\begin{equation}
		\left\{\begin{array}{c}
			\e^{-Q_{00}}+\e^{-\alpha Q_{11}}\le2\\
			\e^{-\alpha Q_{00}}+\e^{-Q_{11}}\le2
		\end{array},\right.\label{eq: noisy 1-bit channel_asym_condition_FT}
	\end{equation}
	and the condition equivalent to the 2nd law:
	\begin{equation}
		\frac{1+\alpha}{2}(Q_{00}+Q_{11})-\log\frac{\e^{\frac{1+\alpha}{2}Q_{00}}+\e^{\frac{1+\alpha}{2}Q_{11}}}{2}\ge0.\label{eq: noisy 1-bit channel_asym_condition_2nd law}
	\end{equation}
	$(Q_{00},Q_{11},\alpha)$ satisfies (\ref{eq: noisy 1-bit channel_asym_condition_FT}), (\ref{eq: noisy 1-bit channel_asym_condition_2nd law}) are plotted in FIG.\ref{fig: coincidence curve}. We see that there is a curve on which the two regions intersect, which justifies the previous result: the set of all realizable, dissipationless $Q(x\to y)$ forms a $|\mathcal{X}|-1=2-1=1$ dimensional submanifold on the boundary.

	\subsection{(Generalized) Zurek's Bound}\label{subsec: (Generalized) Zurek's Bound}
	
	In this subsection, we focus on another limitation of (the probabilistic form of) the 2nd law (\ref{eq: 2nd law probabilistic}), which leads to our derivation of the \textbf{(generalized) Zurek's bound}\cite{kolchinskyGeneralizedZureksBound2023} by using condition (\ref{eq: heat condition_FT}) obtained from the fluctuation theorem.
	
	To illustrate this limitation, consider the following everyday scenario:
	
	\textit{You have 1GB of data (a string of approximately $10^{10}$ zeros and ones) stored on your hard disk. Now, you wish to erase this specific string of bits, resetting all the bits to zeros. What is the minimum amount of energy that must be dissipated as heat to perform this erasure?}
	
	Does the second law (\ref{eq: 2nd law probabilistic}) provide an answer to this question? Unfortunately, no. Since the 2nd law depends on the initial distribution $p_X$, and this scenario does not specify such a distribution, but only an initial state "10110..", the 2nd law teaches us nothing on the minimum energy dissipated, or the minimum heat released during the erasure process.
	
	In some sense, this silence of the 2nd law is justified because, for every given initial state, one could theoretically design a highly specific process (e.g., "flip the 1st bit, keep the 2nd bit, flip the 3rd bit..."), that is dissipationless for that particular input.
	
\begin{figure}
		\centering
		\includegraphics[width=0.28\textwidth]{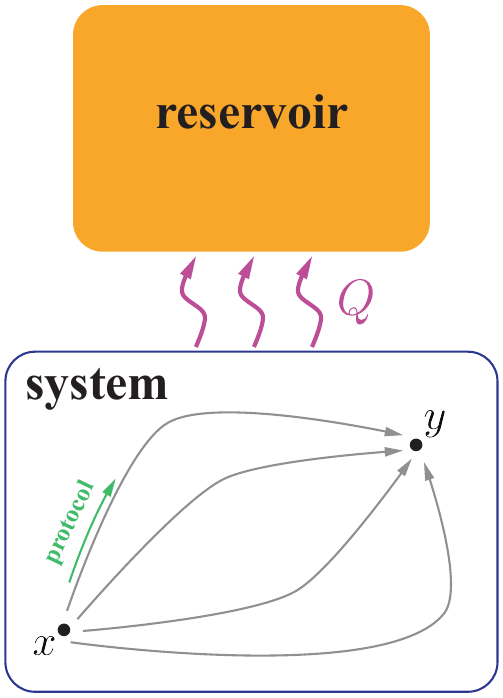}
		\caption{A system coupled with a reservoir is driven by a protocol $\mathcal{P}$.  In different attempts, the system may travel different trajectories. $Q(x\to y)$ is the average heat release over all trajectories from $x$ to $y$.}
		\label{fig: illustration}
	\end{figure}

	However, our processes (machines/protocols) are typically constrained by their complexity and are designed to handle generic inputs rather than highly specific ones. Thus, it is natural to ask: given the "complexity" of the process, is there a lower bound of dissipation required to drive a single state $x\in\mathcal{X}$ to a single state $y\in\mathcal{Y}$ (FIG.\ref{fig: illustration}).
	
	Kolchinsky\cite{kolchinskyGeneralizedZureksBound2023} recently provided a rigorous answer to this question by deriving the so-called "(generalized) Zurek's bound" in the context of the quantum \textbf{two-point measurement} scheme\cite{campisiColloquiumQuantumFluctuation2011}:
	\begin{equation}
		Q(x\to y)\overset{+}{\ge}K(x|y,\mathcal{P})+\log P(y|x),\label{eq: Zurek bound}
	\end{equation}
	where $\mathcal{P}$ denotes the process that we use to drive the system, and $K(x|y,\mathcal{P})$ denotes the \textbf{conditional Kolmogorov complexity} of the initial state $x$, given the final state $y$ and the process $\mathcal{P}$ (see Appendix \ref{app: AIT}). Intuitively, we can regard $K(x|y,\mathcal{P})$ as "the amount of information required to compute $x$ from $y$ and $\mathcal{P}$". We put a "+" on top of "$\ge$" to mean that the inequality holds within a constant additive term.
	
	The "(generalized) Zurek's bound" (\ref{eq: Zurek bound}) is named after Zurek, who first conjectured that the minimal dissipation should be the conditional Kolmogorov complexity $K(x|y)$. Although simple, this bound is incorrect because it overlooks the information encoded in the process $\mathcal{P}$.
	
	Let's write the bound (\ref{eq: Zurek bound}) as
	\begin{equation}
		K(x|y,\mathcal{P})\overset{+}{\le}Q(x\to y)-\log P(y|x)\label{eq: Zurek bound1}
	\end{equation}
	Since
	\begin{equation}
		\sum_x\e^{-(Q(x\to y)-\log P(y|x))}=\sum_xP(y|x)\e^{-Q(x\to y)}\le1,
	\end{equation}
	where the inequality follows from (\ref{eq: heat condition_FT}), we can define a (non-normalized) probability distribution 
	\begin{equation}
		Pr(x|y)\equiv\e^{-(Q(x\to y)-\log P(y|x))}.
	\end{equation}
	The bound to be proven (\ref{eq: Zurek bound1}) becomes
	\begin{equation}
		K(x|y,\mathcal{P})\overset{+}{\le}-\log Pr(x|y).
	\end{equation}
	Since $Pr(\cdot|\cdot)$ can be computed from $P(\cdot|\cdot)$ and $Q(\cdot\to\cdot)$, and both of which can be computed from $\mathcal{P}$, we have
	\begin{equation}
		K(x|y,\mathcal{P})\overset{+}{\le} K(x|y,P,Q)\overset{+}{\le}K(x|y,Pr)\overset{+}{\le}-\log Pr(x|y),
	\end{equation}
	where the last inequality follows from the coding theorem (\ref{app_eq: coding thm}).
	
	Our derivation differs from Kolchinsky's in that our assumption of the fluctuation theorem is not limited to the two-point measurement scheme but holds for many other systems as well. For instance, some quantum processes obey the detailed fluctuation theorem but cannot be described by the two-point measurement scheme\cite{micadeiQuantumFluctuationTheorems2020}.

	\section{Conclusions}\label{sec: Conclusions}
	
	In this article, we derived the condition for realizability of heat released in nondeterministic processes from the differential fluctuation theorem. While this condition was previous derived in \cite{turgutRelationsEntropiesProduced2009} for Hamiltonian systems, our derivation from the fluctuation theorem generalizes it to any system that satisfies a fluctuation theorem.
	
	We demonstrated that this condition for realizability is strictly stronger than the 2nd law: there exist pairs $(P(y|x),Q(x\to y))$ which satisfy the 2nd law but contradict the fluctuation theorems. We also proved the existence of realizable dissipationless heat release function $Q(x\to y)$ given any process $P(y|X)$.
	
	Moreover, using our condition for realizability, a straightforward derivation of the (generalized) Zurek's bound from the fluctuation theorems can be established, thereby expanding the applicability of the bound.

	\section{Ackonwledgements}
	We thank Artemy Kolchinsky for helpful suggestions.

	\bibliographystyle{apsrev4-2}
	\bibliography{references}
	
	\appendix
	\renewcommand{\appendixname}{APPENDIX}

	\section{Algorithmic Information Theory}\label{app: AIT}
	We briefly sketch the notations and results of algorithmic information theory(AIT) that are used when discussing the (generalized) Zurek's bound. For more details, readers can refer to \cite{caludeInformationRandomnessAlgorithmic2002,grunwaldShannonInformationKolmogorov2004}.
	
	AIT is a study of the "complexity" of objects (numbers, sequences, functions, etc.). By the term "complexity" of an object $x$ in a language $U$, or $K_U(x)$ as is the usual notation, we mean the minimal resource required to "describe" $x$ in $U$. Usually, $U$ is assumed to be a (prefix-free) universal programming language (e.g., a universal Turing machine, Lisp, C++, Python, etc.) and $K_U(x)$ is defined as the length of the shortest program $p$ in $U$ that produces $x$ as the output. That is:
	\begin{equation}
		K_U(x)\equiv\min\{l(p):U(p)=x\},\label{app_eq: Kolmogorov0}
	\end{equation}
	where $U$ is the specific universal programming language we select and $p$ is a program in $U$.
	
	We notice from (\ref{app_eq: Kolmogorov0}) that $K_U(x)$ is dependent on the artificially selected universal programming language $U$. Fortunately, we can escape from this subjectivity by noticing that there exists a compiler program $cmpl(U,U')$ between any two universal programming languages $U$ and $U'$. That is, any program in $U'$ can be translated to a program in $U$ by just inserting $cmpl(U,U')$ before the program. Therefore:
	\begin{equation}
		K_U(x)-K_{U'}(x)\leq l(cmpl(U,U')),\label{compiler}
	\end{equation}
	and vice versa, which implies:
	\begin{equation}
		K_U(x)\overset{+}{=}K_{U'}(x),\label{invariance}
	\end{equation}
	so definition (\ref{app_eq: Kolmogorov0}) does captures some objective "complexity" of $x$ that is irrelevant to the language, and there's nothing lost in writing $K(x)$ instead of $K_U(x)$.
	
	The relative Kolmogorov complexity $K(x|y)$ is similarly defined as $K(x)$, except for that the program $p$ can access to a prior information given by $y$ anytime during computation. To be specific:
	\begin{equation}
		K_U(x|y)\equiv\min\{l(p):U(p,y)=x\}.\label{eq:kolmogorov1}
	\end{equation}
	Similarly, we may write $K(x|y)$ instead of $K_U(x|y)$. It is easy to see that $K(x)\overset{+}{=}K(x|\epsilon)$, where $\epsilon$ is the empty string.
	
	The following statements of Kolmogorov complexity hold:
	\begin{enumerate}
		\item \textbf{Kraft's inequality}: for every $y$,
		\begin{equation}
			\sum_x\e^{-K(x|y)}\le1.\label{app_eq: Kraft ineq}
		\end{equation}
		\item For every $x$, $y$ and $z$,
		\begin{equation}
			K(x|z)\overset{+}{\le}K(x,y|z)\overset{+}{\le}K(x|y,z)+K(y|z).\label{kolmogorov_inequality} 
		\end{equation}
		\item\textbf{The coding theorem}: for every (computable) conditional distribution $P(x|y)$ such that $\sum_xP(x|y)\leq 1$ for every $y$, 
		\begin{equation}
			K(x|y,P)\overset{+}{\leq}-\log P(x|y).\label{app_eq: coding thm}
		\end{equation}
		Note that $P(x|y)$ may be non-normalized, although we refer to it as a "distribution".
	\end{enumerate}
	
	There is one more point to be noted: we haven't specify the size of the alphabet for our universal programming language $U$ in the above. Suppose two universal programming language $A$ and $B$, running on two alphabets of $a$ symbols and $b$ symbols respectively. It is easy to see that $K_A(x)/\log a\overset{+}{=}K_B(x)/\log b$. Since $\log$ is assumed to be the natural logarithm in this article, to justify formulas (\ref{app_eq: Kraft ineq}) and (\ref{app_eq: coding thm}), we have to require that $U$ runs on an alpabet of $\e\approx2.718$ symbols (thus the unit for $K(x)$ is \textit{nat}), or to write $(\log 2)K(x)$ instead of $K(x)$ if $U$ runs on a 2-symbol alphabet (thus the unit for $K(x)$ is \textit{bit}).
	
\end{document}